\documentclass[twocolumn,9pt]{extarticle}
\usepackage{graphicx}
\usepackage[small]{caption}
\usepackage{color}
\usepackage[all]{xy}

\usepackage{amsmath}
\usepackage{epsfig}
\usepackage{txfonts}

\begin{document}

\title{Impact of noise on a dynamical system: prediction and uncertainties from a swarm-optimized neural network}
\author{C.~H.~L\'opez-Caraballo, J.~A.~Lazz\'us, I.~Salfate, P.~Rojas, M.~Rivera and L.~Palma-Chilla\\
\textsf{\small Departamento de F\'isica y Astronom\'ia, Universidad de La Serena, Avda. J. Cisternas 1200, La Serena, Chile.} \\ \\
\textsf{\small Correspondence should be addressed to L\'opez-Caraballo C. H.; clopez@dfuls.cl} \\
\textsf{\small Published in {\it Computational Intelligence and Neuroscience} (2015). (http://dx.doi.org/10.1155/2012/351836) }}

\date{}
\maketitle

\begin{abstract}
\noindent In this study, an artificial neural network (ANN) based on particle swarm optimization (PSO) was developed for the time series prediction. The hybrid ANN+PSO algorithm was applied on Mackey--Glass chaotic time series in the short-term $x(t+6)$. The performance prediction was evaluated and compared with another studies available in the literature. Also, we presented properties of the dynamical system via the study of chaotic behaviour obtained from the predicted time series. Next, the hybrid ANN+PSO algorithm was complemented with a Gaussian stochastic procedure (called {\it stochastic} hybrid ANN+PSO) in order to obtain a new estimator of the predictions, which also allowed us to compute uncertainties of predictions for noisy Mackey--Glass chaotic time series. Thus, we studied the impact of noise for several cases with a white noise level ($\sigma_{N}$) from 0.01 to 0.1.
\end{abstract}

\section{Introduction}

Currently, the prediction of time series has played an important role in many science fields of practical application as engineering, biology, physics, meteorology, etc. In particular, and due to their dynamical properties, the analysis and prediction of chaotic time series have been of interest for the science community. In general, the chaotic time series are usually modeled by delay-differential equations; standard examples are the Mackey--Glass system \cite{mackey1977}, or the Ikeda equation \cite{ikeda1979} (for more examples see \cite{bez2001}).
Also, many methods have been used in the chaotic time series analysis \cite{hamilton1994}.
However, in the last decades different types of artificial neural networks (ANN) have been widely used for forecasting of chaotic time series, for example, back-propagation algorithm \cite{kuruna2006}, radial basic function \cite{chng1996}, recurrent network \cite{zhang2000}, etc.

On the other hand, the analysis of real-life time series requires of taking into account the error propagation of input uncertainties. The observed data could be contaminated for different instrumental noise types as white noise or proportional to signal (the latter mainly arises from instrumental calibration).
In modeling of chaotic time series, the impact of noise can be treated as errors-invariable problem where the noise is propagated into the prediction model. In the literature, the noisy impact on chaotic time series prediction has been barely considered. We can found studies where the algorithms were tested from a theoretical point of view (for example, see \cite{girard2003,sapan2009,haykin1998,li2012,han2013}), and works where the implementation was applied on real-life time series (for example, see \cite{leung2001,sheng2012,sapan2009}).
In addition, some authors have proposed a modification to the standard methods in order to improve the performance prediction in presence of noise \cite{sheng2012,sapan2009}.

In this work, we used the Mackey--Glass chaotic time series in order to study the short-term prediction ($x(t+6)$) with an artificial neural network optimized with a particle swarm algorithm (ANN+PSO).
The method was applied on noiseless and noisy chaotic time series.
In order to carry out the error propagation of the input noise, this hybrid algorithm was complemented with a Gaussian stochastic procedure to compute a new estimator of the predictions and their uncertainties. 
Note that ANNs have been used in combination with PSO in several applications. Principally, these applications include: feedforward neural network training \cite{Han2011,WeiChang2013,Zhang,Grimaldi}, design of recurrent neural networks \cite{Juang2004}, design of radial basis function networks \cite{Huang2007}, and neural network control for nonlinear processes \cite{Song2007}. In addition, there are several current versions of PSO available in the literature (for example, see the following reviews \cite{Banks2007,Banks2008,Thangaraj2011}), but our application uses a standard PSO with inertial weight \cite{Shi1998MPSO}. 
In this point, the use of a PSO with inertial weight is based on the following reasons: 1) this vesion of PSO is easy to understand and implement due to its simple concept and learning strategy; 2) as pointed out in \cite{Eberhart2000}, the PSO with inertia weight \cite{Shi1998MPSO} and PSO with constriction factor \cite{Clerc2002} are mathematically equivalent, and PSO with constriction factor can be considered as a special case of PSO with inertia weight \cite{Banks2007,Eberhart2000} (note that this equivalence can be applied to other improved PSO algorithms that include a varying the inertia weight schedule); 3) inertia weight PSO algorithm is quite stable to population changes \cite{Banks2008}; 4) the advantages and disadvantages of variants of PSO depend on the problem to solve \cite{Banks2007,Banks2008,Thangaraj2011}; 5) as a first approach for study of noise effect on dynamical systems using an ANN combined with inertia weight PSO algorithm, so the present study may motivate and help the researchers working in the field of evolutionary algorithms to develop new hybrid models or to apply other existing PSO models to solve this problem. 
To the best of the authors's knowledge, there is no application for forecasting of noisy chaotic time series such as the one presented here, using a hybrid method that combined ANN with PSO algorithm.

Organization of this paper is as follows. In Section~\ref{sec:ANN+PSO} we present a detailed description of the hybrid ANN+PSO method. Section~\ref{sec:noiseless} and Section~4 present the simulation, algorithm implementation and the principal results obtained for the forecasting of noiseless chaotic time series and noisy time series, respectively. Finally, conclusions are given in Section~5.

\section{Hybrid ANN+PSO algorithm}
\label{sec:ANN+PSO}
Artificial neural networks (ANN) are similar to biological neural networks in performing functions collectively and in parallel using connection nodes.  Thus, ANN are a family of statistical learning algorithms biologically inspired. 

In this study, we consider one of the most successful and frequently used types of neural networks: a multilayer feed-forward neural network with a backpropagation learning algorithm (gradient descent error). This ANN was implemented replacing standard backpropagation with particle swarm optimization (PSO).

PSO is a population-based optimization tool, where the system is initialized with a population of random particles and the algorithm searches for optima by updating generations \cite{lazzusj2009}. In each iteration, the velocity of each particle $j$ is calculated according to the following formula \cite{lazzus2010}:

\begin{equation}
v_{j}^{k+1}=\omega v_{j}^{k}+c_{1}r_{1}\left(\psi_{j}^{k}-s_{j}^{k}\right)+c_{2}r_{2}\left(\psi_{g}^{k}-s_{j}^{k}\right)
\label{eq2}
\end{equation}

\noindent where $s$ and $v$ denote a particle position and its corresponding velocity in a search space, respectively. $k$ is the current step number, $\omega$ is the inertia weight, $c_{1}$ and $c_{2}$ are the acceleration constants, and $r_{1}$, $r_{2}$ are elements from two random sequences in the range (0,1). $s_{j}^{k}$  is the current position of the particle, $\psi_{j}^{k}$ is the best one of the solutions that this particle has reached, and $\psi_{g}$ is the best solutions that all the particles have reached. In general, the value of each component in $v$ can be clamped to the range [$-v_{\rm max}$, $+v_{\rm max}$] control excessive roaming of particles outside the search space \cite{lazzusj2009,lazzus2010}. After calculating the velocity, the new position of each particle is:

\begin{equation}
s_{j}^{k+1}=s_{j}^{k}+v_{j}^{k+1}
\label{eq3}
\end{equation}

The procedure to calculating the output values, using the input values are described in detail in \cite{lazzus2009}.

The net inputs ($N$) are calculated for the hidden neurons coming from the inputs neurons. In the case of a neuron in the hidden layer is has:

\begin{equation}
N_{i}^{h}=\sum_{i}^{n}w_{i,j}^{h}p_{i}+b_{i,j}^{h}
\label{eq:ann_neurons}
\end{equation}

\noindent where $p_{i}$ is the vector of the inputs of the training, $w_{i,j}^{h}$ is the weight of the connection among the input neurons with the hidden layer $h$, and the term $b_{i,j}^{h}$ corresponds to the bias of the neuron of the hidden layer $h$, reached in its activation. The PSO algorithm is very different then any of the traditional methods of training \cite{lazzusj2009}. Each neuron contains a position and velocity. The position corresponds to the weight of a neuron $\left(s_{i}^{k}\rightarrow w_{i,j}^{h}\right)$. The velocity is used to update the weight $\left(v_{i}^{k+1}\rightarrow w'_{i,j}\right)$. Starting from these inputs, the outputs ($y_{i}$) of the hidden neurons are calculated, using a transfer function $f^{h}$ associated with the neurons of this layer:
\begin{equation}
y_{i}=f^{h}\left(\sum_{i}^{n}w_{i,j}^{h}p_{i}+b_{i,j}^{h}\right)
\label{eq:ann_out}
\end{equation}

The transfer functions $f^{h}$ can be linear or non-linear. We used one hidden layer with $f^{h}_{i}$ as a tangent hyperbolic function ({\it tansing}) and $f^{h}_{j}$ as a  linear function in the output layer.

\begin{equation}
f(N_{i})=\frac{e^{N_i}-e^{-N_i}}{e^{N_i}+e^{-N_i}}
\label{eq6}
\end{equation}
%
\begin{center}
\begin{table}[t!!!!]
\caption{Parameters used in the hybrid {\sc ANN+PSO} algorithm.}
\centering
\begin{tabular}{l c}
\hline 
& \\ [-2  ex]
 \multicolumn{2}{c}{ANN}        \\ [0.5 ex]
\hline
   & \\ [-2.3  ex]
NN-type & feed-forward           \\               
Number of hidden layers & 1      \\                
Transfer function & {\it tansig} \\     
Number of iterations & 1500      \\            
Normalization range & [--1, 1]   \\      
Weight range & [--100, 100]      \\      
Bias range & [--10, 10]          \\       
Minimun error & 1e--3            \\ [0.5 ex]
\hline 
   & \\ [-2  ex]
 \multicolumn{2}{c}{PSO}  \\ [0.5 ex]
\hline 
   & \\ [-2.3  ex]
Number of particles in swarm ($N_{\rm part}$) & 50\\  
Number of iterations ($k_{\rm max}$) & 1500\\          
Cognitive component ($c_{1}$) & 1.494\\                
Social component ($c_{2}$) & 1.494\\                   
Maximum velocity ($v_{\rm max}$) & 12\\                
Minimum inertia weight ($\omega_{\rm min}$) & 0.5\\    
Maximum inertia weight ($\omega_{\rm max}$) & 0.7\\   %
Objective function & RMSE\\    
\hline         
\end{tabular}          
\label{tab:input_ANN_PSO}            
\end{table}             
\end{center}
All the neurons of the ANN have an associated activation value for a give input pattern, and the algorithm continues finding the error that is presented for each neuron, except those of the input layer. After finding the output values, the weights of all layers of the network are actualized $w_{i,j}\rightarrow w'_{i,j}$ by PSO, using eqs. (\ref{eq2} and \ref{eq3})\cite{lazzus2010}. The velocity is used to control how much the position is updated. On each step, PSO compares each weight using the data set. The network with the highest fitness is considered the global best. The other weights are updated based on the global best network rather than on their personal error or fitness \cite{lazzusj2009}. In this article, we used the mean square error (MSE) to determine network fitness for the entire training set:
\begin{equation}
{\rm MSE}=\frac{\sum_{i=1}^{n}\left(Y_{i}^{\rm true}-Y_{i}^{\rm calc}\right)^{2}}{n}
\label{eq7}
\end{equation}
where $Y_{i}^{\rm true}$ is the real data and $Y_{i}^{\rm calc}$ is the calculated output value obtained from the normalized output ($y_{i}$) of the network. This process was repeated for the total number of patterns in the training set. For a successful process the objective of the algorithm is to modernize all the weights minimizing the total root mean squared error (RMSE):
\begin{equation}
{\rm RMSE}=\sqrt{\rm MSE}
\label{eq:rmse}
\end{equation}
\begin{equation}
\varepsilon=\min ({\rm RMSE}).
\label{eq9}
\end{equation}

\begin{figure}[t!!!]
\begin{center}
\includegraphics[width= 7.9 cm]{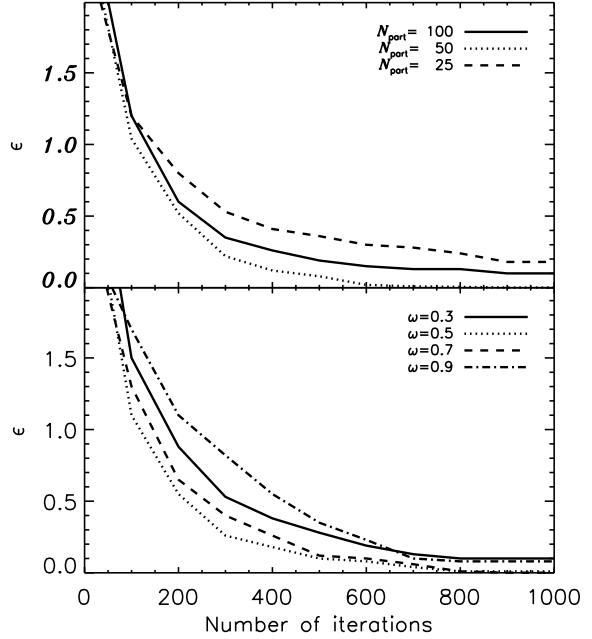}
\caption{Illustration of the behaviour of some parameters of the ANN+PSO against the number of iterations. The top and bottom panel correspond to the number of particles in the swarm ($N_{part}$) and the inertia weight ($\omega$), respectively.}
\label{fig:figure_convegency}
\end{center}
\end{figure}

In PSO, the inertial weight $\omega$, the constant $c_{1}$ and $c_{2}$, the number of particles $N_{\rm part}$ and the maximum speed of particle summary the parameters to synchronize for their application in a given problem. 
Then, an exhaustive trial-and-error procedure was applied for tuning the PSO+ANN parameters. Firstly, the effect of population $N_{\rm part}$ is analyzed for values of 25 to 100 individuals in the swarm. 
For other applications, some authors have shown that a larger swarm increases the number of function evaluations to converge to an error limit \cite{Ratnaweera2004}.
In addition, Shi and Eberhart \cite{Shi1998} illustrated that the population size has hardly any effect on the performance of a swarm algorithm. The top panel of Figure~\ref{fig:figure_convegency} shows that the best population to solve the problem is of 50 individuals.  
Next, the effect of $\omega$ is analyzed for values of 0.1 to 0.9. The bottom panel of Figure~\ref{fig:figure_convegency} shows the values of $\omega$ that favoured the search of the particles and accelerated the convergence. This figure shows that for a linearly decreasing inertia weight starting at 0.7 and ending at 0.5, the PSO+ANN presents a good convergence. In other aspect, a usual choice for the acceleration coefficients $c_1$ and $c_2$ is $c_1$\,=\,$c_2$ \cite{Ratnaweera2004}. The effect of variation of constants was evaluated for the commonly used values of $c_1$ and $c_2$ such as 1.49, and 2.00 \cite{Ratnaweera2004,Shi1998}. For this analysis, $c_1$\,=\,$c_2$\,=\,1.49 presents a better convergence than other values.
Table~1 shows the selected parameters for this hybrid algorithm.

The step-to-step approach of PSO+ANN can be summarized as:

\noindent \textbf{Step 1:} Initialize the positions (weights and biases) and velocities of a group of particles randomly. The particles represents the weight vectors of ANN, including biases. The dimension of the search space is therefore the total number of weights and biases.

\noindent \textbf{Step 2:} The ANN is trained using the initial particles position in PSO. The learning error produced from ANN network can be treated as particles fitness value according to initial weight and bias. The current best fitness achieved by particle $j$ is set as $\psi_{j}^{k}$. The $\psi_{j}^{k}$ with best value is set as $\psi_{g}$ and this value is stored.

\noindent \textbf{Step 3:} Evaluate the desired optimization fitness function (Eq. 7) over a given data set.

\noindent \textbf{Step 4:} Compare the evaluated fitness value of each particle ($F_j$) with its   value. If $F_j < \psi_{j}^{k}$  then $\psi_{j}^{k}=s_{j}^{k}$ is the coordinates corresponding to best particle so far.

\noindent \textbf{Step 5:} The objective function value is calculated for new positions of each particle. If a better position is achieved by an agent, $\psi_{j}^{k}$ value is replaced by the current value. As in Step 1, $\psi_{g}$ value is selected among $\psi_{j}^{k}$ values. If the new $\psi_{g}$ value is better than previous value, it is replaced by the current $\psi_{g}$ value and this value is stored. if $F_j < \psi_{g}$ then $\psi_{g}=s_{j}^{k}$ is the particle having the overall best fitness over all particles in the swarm.

\noindent \textbf{Step 6:} The learning error at current epoch will be reduced by changing the particles position, which will update the weight and bias of the network. Change the velocity and location of the particle according to movement equations (Eqs. 1 and 2). The new sets of positions (weights and biases) are produced by adding the calculated velocity value to the current position value. Then, the new sets of positions are used to produce new learning error in ANN.

\noindent \textbf{Step 7:} This process is repeated until the stopping conditions either minimum learning error or maximum number of iteration are met, then stop; otherwise Loop to Step 3 until convergence. 

\noindent \textbf{Step 8:} The optimum weight and biases for ANN model are obtained by PSO. Best training process is obtained for ANN.

In our time series analysis, if the input noise level contribution is available, the RMSE in the training phase shall be computed as follow:
\begin{equation}
{\rm RMSE}= \sqrt{ \frac{1}{n}\sum_{i=1}^{n} \frac{\left(Y^{\rm cal}_{i} - Y^{\rm true}_{i} \right)^2}{\sigma_{N,i}^2}}
\label{eq:rmse_std}
\end{equation}
where $\sigma_{N,i}$ is the noise level of each $i$--element. Note that $\sigma_{N,i} = \sigma_{N}$ for a white noise assumption.

Henceforth, we refer as {\it the standard ANN+PSO} to the hybrid ANN+PSO defined above.

\subsection{The {\it Stochastic} {\sc ANN+PSO}}
\label{sec:ANN+PSO_stochastic}
Up to now, the standard ANN+PSO is not developed to carry out the error propagation of the input noise level contribution.
Nevertheless, once the {\it standard} {\sc ANN+PSO} has been executed and has provided the optimal topology, we can apply an additional method in order to compute uncertainty of the prediction.

Note that once the topology is established (number of hidden layer, neurons in each hidden layer, transfer functions $f^{h}$, and weights and biases ($w^h_{i,j}$ and $b^h_{i,j}$)), the neural network acts as a function (called {\it function ANN}) whose output only depends on the input vector (see, Eq. 4).
The idea is to generate simulations from the input data ($d_i\equiv d(t)$) via Gaussian random number generator in order to propagate the intrinsic data noise through the {\it function ANN}.

For each $i$--element of the input time series we generate $k$--simulations as:
\begin{equation}
d_{i,k} = d_i + GR_k(\sigma_{N,i})
\label{eq:sim_d}
\end{equation}
\noindent where the input noise level $\sigma_{N,i}$ is known. $GR(\sigma_{N,i})$ is a random number generator following a Gaussian distribution with mean zero and standard deviation equal to $\sigma_{N,i}^2$.

Finally, and for the $i$-th element, each $k$ input data $d_{i,k}$ provides an output $y_{i,k}$. These $y_{i,k}$ are used in the estimation of a new estimator of prediction ($\hat{y}_i$) and an error on the prediction ($\sigma_{\hat{y}}$) as follows:
\begin{equation}
\hat{y}_i = <y_{i,k}> \ \ {\rm and} \ \ \sigma_{\hat{y}}= <y_{i,k}^2>^{1/2}.
\end{equation}

\section{Noiseless chaotic time series prediction}
\label{sec:noiseless}
We computed the chaotic time series from the Mackey--Glass time-delay differential system \cite{mackey1977,farmer1982}, which is described as follow:
\begin{equation}
\frac{dx}{dt}=\beta x(t)+\frac{\alpha x(t-\tau)}{1+x(t-\tau)^{10}}
\label{eq10}
\end{equation}
\noindent where $x$ (unitless) is the series in the time $t$, and $\tau$ the time delay.
Here, we assumed $\alpha=0.2$, $\beta=0.1$ and $x(0)=1.2$. Note that, if $\tau\geq$17 the time series shows a chaotic behaviour \cite{mirzaee2009,farmer1982}. The nominal Mackey--Glass time series is obtained from numerical integration by a fourth order Runge--Kutta method. This series was computed with a time sampling of 1 second. Thus, $x(t)$ is derived for $0\leq t \leq t_{\rm h}$ with $x(t)=0$ for $t<0$, where $t_{\rm h}$ is the time horizon considered. 

Mackey--Glass chaotic time series with $\tau=$\,17 is considered as the nominal case $x^{\rm Noiseless}$ (without noise contribution). Here, we generate two thousand data points ($t_{\rm h}=2000$).

From this data set, the input is created as a vector using $d$ points of the time series spaced $\Delta$ apart, i.e., ${\bf x}(t)=\left[x(t), x(t+\Delta),\cdots, x(t+(d-1)\Delta)\right]$. The output is generated with the value $x(t+T)$.

According to the standard analysis of the Mackey--Glass chaotic time series, we consider four non consecutive points in the chaotic time series in order to predict the short-term $x(t+6)$:
\begin{equation}
x(t+6)=F\left[x(t), x(t-6), x(t-12), x(t-18)\right]
\end{equation}
where this standard test assumes $d=4$ and $\Delta=T=6$ \cite{chng1996,mirzaee2009}.

%
\begin{figure}[h!!!]
\begin{center}
\includegraphics[trim= 0cm 1.cm 0cm 0.5cm,clip=true,width= 6.9 cm]{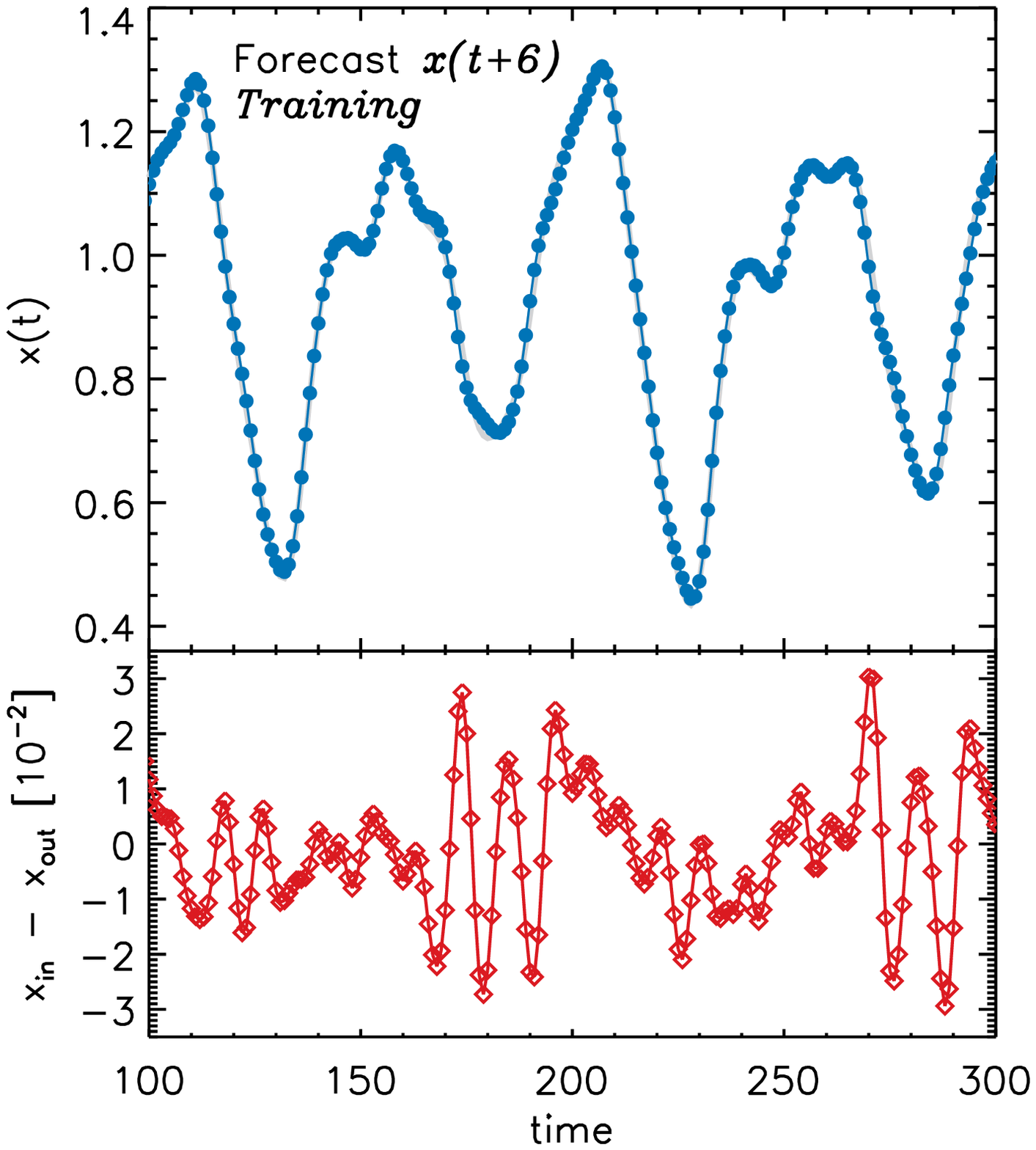}
\includegraphics[trim= 0cm 1.cm 0cm 0.5cm,clip=true,width= 6.9 cm]{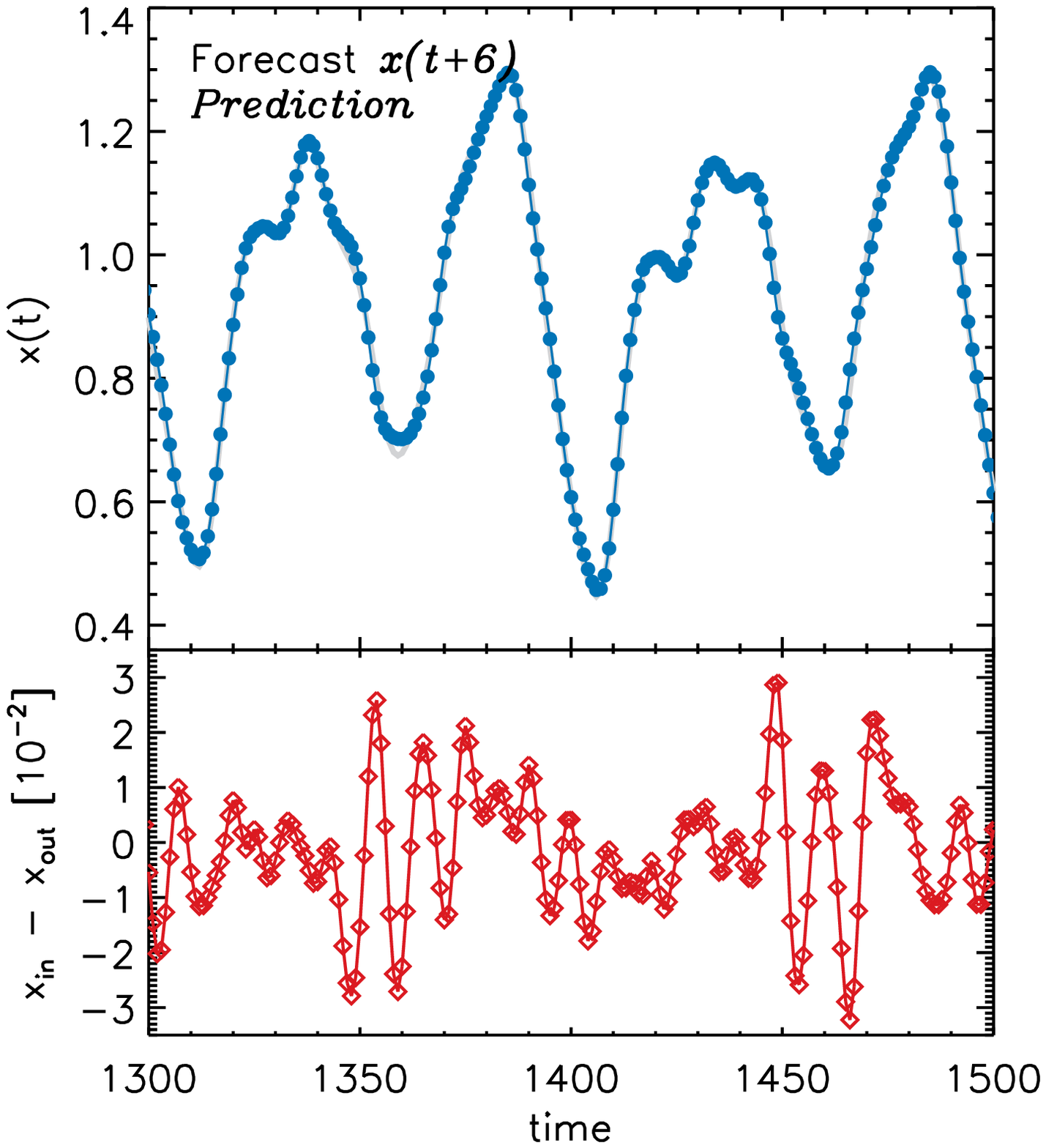}
\caption{Performance of ANN+PSO method on the Mackey--Glass chaotic time series (noiseless). The top and bottom panel show the {\it training} and {\it prediction} performance for the short-term $x(t+6)$ analysis, respectively. The grey and blue lines correspond to the input ($x_{\rm in}$) and output ($x_{\rm out}$) data. The red line with diamond draws the difference between the input and output data (in a factor of $10^{-2}$).}
\label{fig:figure1}
\end{center}
\end{figure}

For this input, the first thousand data were used for learning ({\it training}) while the others were used for the prediction validation ({\it prediction}).
In the ANN+PSO implementation on the nominal case, the optimum value of $N_{\rm HL}$ found was six, i. e., the architecture is described as 4-6-1.

Figure~\ref{fig:figure1} present a comparison between recorded and predicted values of the Mackey--Glass time series for the training and prediction phases. This figure shows that for training and validation phases, the nominal and reconstructed values are in total agreement.
In fact, and for training, we computed a remainder average, $<x_{\rm in} - x_{\rm out}>$, of $-1.4\times10^{-3}$ and a remainder maximum, max$\{|x_{\rm in} - x_{\rm out}|\}$, of $3.20\times10^{-2}$. Similar results are obtained for the prediction phase; with a maximum of $3.22\times10^{-2}$ and an average of $-1.5\times10^{-3}$.


\begin{center}
\begin{table}[t!!!!]
\caption{Root mean squared error (RMSE) reported for different methods in the Mackey--Glass chaotic time series analysis.}
\centering
\begin{tabular}{l c}
\hline
 Method & RMSE$_{x(t+6)}$ \\
\hline
\multicolumn{2}{c}{}  \\ [-2.0 ex]
Linear model \cite{lazzus2014}& 0.5503  \\         
Conjugate gradient ANN \cite{Zhao2014} & 0.2296 \\                                                               
Product operator T-norm \cite{Wang1992} & 0.0907 \\
Fuzzy system \cite{Lee1994} & 0.0816 \\
Cascade correlation NN \cite{Chen2005}& 0.0624 \\    
Genetic algorithm and fuzzy system \cite{Kim1997} & 0.0490 \\                                                          
Backpropagation NN \cite{lazzus2014} & 0.0262 \\                                                                                                                            
Linguistic Model (20 rules ) \cite{Qin2014}& 0.0256 \\
K-Nearest Neighbor \cite{Yen2006} & 0.0194 \\
This work & 0.0138  \\                                                           
\hline                                                                                                                                                   
\end{tabular}
\label{tab:noiseless_rmse}
\end{table}
\end{center}

Table~\ref{tab:noiseless_rmse} shows the RMSE (for short term prediction of  Mackey--Glass chaotic time series) from different computational methods obtained from literature, for example, the Back-propagation NN \cite{lazzus2014}, the conjugate gradient ANN \cite{Zhao2014}, the product operator T-norm \cite{Wang1992}, the fuzzy system \cite{Lee1994}, etc (see references in Table~\ref{tab:noiseless_rmse}).
In the ANN+PSO configuration used here, the RMSE\,=\,0.014 indicates that the performance prediction is in good agreement with other methods. Clearly, the inclusion of the PSO approach allows us to improve methods based on ANN without PSO as, for example, the conjugate gradient ANN (RMSE\,=\,0.229) and the back--propagation NN (RMSE\,=\,0.026).

\subsection{Chaotic behaviour}
\begin{figure}[t!!!!!!!]
\begin{center}
\includegraphics[width= 6.9 cm]{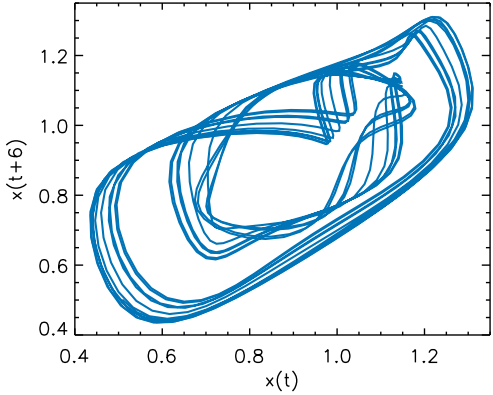}
\caption{Chaotic attractor for the Mackey--Glass noiseless chaotic time series ($\tau=17$).}
\label{fig:figure3}
\end{center}
\end{figure}

As the Mackey--Glass time series without noise, it is a known system, is possible to compare the ability of ANN+PSO method of reproducing its chaotic behavior. Figure \ref{fig:figure3} shows a representation of the chaotic attractor studied from Mackey--Glass time series. This Figure shows that with $\tau=17$ the system operates in a high-dimensional regime. The Mackey--Glass system is infinite dimensional (because it is a time-delay equation) and, thus, has an infinite number of Lyapunov exponents ($\lambda_{i}$) \cite{farmer1982}. The Lyapunov exponents of dynamical systems are one of a number of invariants that characterize the attractors of the system in a fundamental way \cite{brown1997}. Table~3 shows a comparison of the first four largest Lyapunov exponents of the Mackey--Glass system reported in \cite{farmer1982}, with the Lyapunov exponents obtained for the ANN+PSO method for $\tau=$17.

\begin{center}
\begin{table}[t!!!!]
\caption{Lyapunov exponents reported in Farmer~\cite{farmer1982} versus calculated for the ANN+PSO method.}
\centering
\begin{tabular}{c c}
\hline
\multicolumn{2}{c}{}  \\ [-2.0 ex]
$\lambda_{i}$  & \multicolumn{1}{c}{$\lambda_{i, {\rm ANN+PSO}}$}\\ [0.5 ex]
\hline
0.00860 & 0.00900    \\   
0.00100 & 0.00132    \\                                
--0.03950 & --0.04100\\     
--0.05050 & --0.05000\\ 
\hline
\end{tabular}
\label{tab:par_lyapunov}
\end{table}
\end{center}

An approach to determining an appropriate cutoff value for the number of exponents can be related to the Lyapunov dimension \cite{brown1997}. This idea was originally explored by Kaplan and York \cite{kaplan1979}. Thus, Kaplan and York conjecture that this dimension ($D_{\rm KY}$) is equal to the information dimension \cite{kostelich1989}. In our case $D_{\rm KY}$ is compute as 2.10. Note that in Farmer~\cite{farmer1982}, authors reported a fractal dimension $D_{\rm F}$\,=\,2.13, and a Lyapunov dimension calculated by the Kaplan and York conjecture of $D_{\rm KY}$\,=\,2.10.

\section{Noisy chaotic time series prediction}
\label{sec:noisy}
In the previous section, the ANN+PSO has proven to be an efficient method to the prediction of chaotic time series. Nevertheless, up to now effects of noise on the hybrid ANN+PSO implementation have not been studied.

In order to study the impact of noise on chaotic series time prediction, we constructed the noisy time series as the contribution of a noise level on the nominal case without noise.
The Mackey--Glass noisy chaotic time series, $x_{i} \equiv x(t)$, is generated as:
\begin{equation}
x_{\rm i} = x^{\rm Noiseless}_{\rm i} + \eta_{\rm i} 
\label{eq:sim_dat_mg}
\end{equation}
\noindent where $\eta_{\rm i}$ is the particular contribution of noise on the $i$--element. It is estimated as $\eta_{\rm i} = GR(\sigma_{N,i})$, with $GR(\sigma_{N,i})$ a Gaussian random number generator. 

\begin{figure}[t!!!!!]
\begin{center}
\includegraphics[trim= 1cm 0.1cm 0cm 0.cm,clip=true,width= 8.0 cm]{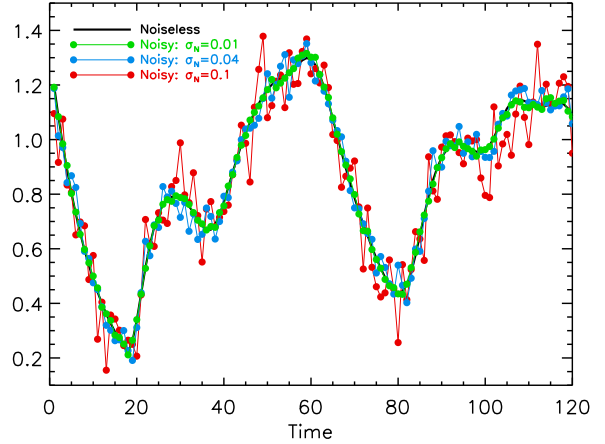}
\caption{Mackey--Glass chaotic time series considered in this work ($\tau=$\,17). The black solid line shows the noiseless case (nominal case). The green, blue and red lines correspond to the Mackey--Glass noisy time series with a white noise level ($\sigma_{N}$) contribution of 0.01, 0.04 and 0.1, respectively.}
\label{fig:MG_noisydata}
\end{center}
\end{figure}

Note that $\sigma_{N,i}^2$ corresponds to the noise level considered. 
Here, we assume that the original data are effected by a white noise, i. e., the noise level is the same in each $i$--element, $\sigma_{N,i} = \sigma_{N}$~\footnote{In order to clarify, although the noise level $\sigma_{N}$ is the same in each time the noise contribution $\eta_{i}$ is not the same (the latter depend on the Gaussian random number generator).}.
Different white noise levels are considered: $\sigma_{N}=0.01$, $\sigma_{N}=0.04$, $\sigma_{N}=0.06$, $\sigma_{N}=0.08$ and $\sigma_{N}=0.1$. 
These values are nearly related with the 1\,\%, 4\,\%, 6\,\%, 9\,\% and 11\,\% of the pick--to--pick amplitude  of nominal case ($ \sim \ x^{Noiseless}_{\rm max} - x^{Noiseless}_{\rm min}$). Figure~\ref{fig:MG_noisydata} shows the noisy chaotic time series for $\sigma_{N}$ equal to 0.01 (green), 0.04 (blue) and 0.1 (red). As expected, the noisy time series with $\sigma_{N}=$\,0.01 is the closest to the nominal case. However, the cases with $\sigma_{N}=$\,0.04 and $\sigma_{N}=$\,0.1 show a slightly more modified shape from the noiseless case, in particular with $\sigma_{N}=$\,0.1.

\subsection{Noise effect on ANN+PSO}
\label{sec:noisy_impact}
The {\it standard} {\sc ANN+PSO} is applied on our noisy time series, which provides the optimum topology and the $y_i$ prediction.
Then, the {\it stochastic} {\sc ANN+PSO} is run in order to obtain a new prediction estimator $\hat{y}_i$ and the uncertainty of the prediction ($\sigma_{\hat{y}_i}$).

\begin{figure}[t!!!!!]
\begin{center}
\includegraphics[width= 8.3 cm]{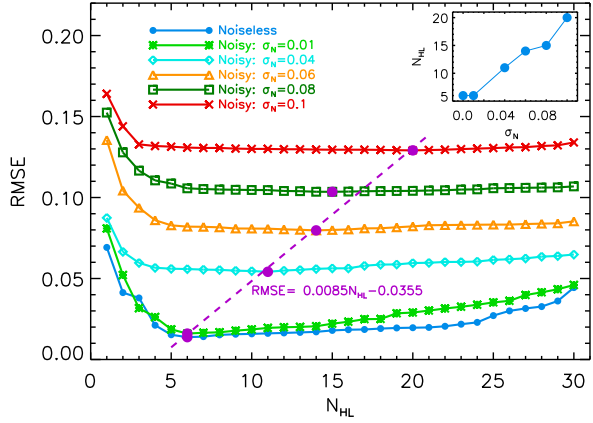}
\caption{Impact of the noise on the architecture.}
\label{fig:MG_output_rmse_vs_nn}
\end{center}
\end{figure}

{\it Impact on architecture.} For each noisy time series, and in the {\it standard} {ANN+PSO} implementation, we carry out a detailed study of the architecture characterization. 
In the determination of the optimum $N_{\rm HL}$, the RMSE is computed for different number of neurons in the hidden layer (from two up to thirty), which are presented in Figure~\ref{fig:MG_output_rmse_vs_nn}.
For each series, the optimum $N_{\rm HL}$ is obtained when the RMSE reachs a minimum.
As expected, the characterization of the architecture is strongly related  with the noise level in the input data. 
In lower noise (as 0.01) the optimum $N_{\rm HL}$ is clearly identified from Figure~\ref{fig:MG_output_rmse_vs_nn}; in contrast, in the most contaminated case ($\sigma_{N}$\,=\,0.1) the selection depends on fourth decimal of the RMSE (0.1292, 0.1291 and 0.1293 for 19, 20 and 21 neurons in the hidden layer, respectively).
The RMSE and the $N_{\rm HL}$ optimum are presented in Table~4.
Using these values, and according to the trend seen in Figure~\ref{fig:MG_output_rmse_vs_nn}, we fit a lineal model, which provides a correlation with a slope of 0.0085.
Although the $N_{\rm HL}$ for $\sigma_{N}$\,=\,0.08 is not well characterized for this model, we can find a clear lineal correlation between the RMSE and the $N_{\rm HL}$ for different noise levels.
In this context, and as illustration, in the overplot  (in top-right side of Figure~\ref{fig:MG_output_rmse_vs_nn}) we show the relation of the $N_{\rm HL}$ and the noise level, whose the best lineal fit model is $N_{\rm HL}$\,=\,146\,$\sigma_{N}$+\,4.7.
Therefore the impact of noise on the architecture of this hybrid neural network, for contributions lower than 0.1, can be characterized by a lineal correlations of the RMSE with the $N_{\rm HL}$, and the $N_{\rm HL}$ with the input noise $\sigma_{N}$.
\begin{center}
\begin{table}[t!!!!]
\caption{Parameters used in the evaluation of the prediction performance of the {\it Standard} and {\it Stochastic} {\sc ANN+PSO} approach.}
\centering
\begin{tabular}{l c c c}
\hline
 & $N_{\rm HL}$ & RMSE & $\xi$ \\
\hline
Noiseless          & 6& 0.0138&  1  \\
$\sigma_{N}=0.01$  & 6& 0.016 &1.2\\
$\sigma_{N}=0.04$  & 11& 0.054&3.9\\
$\sigma_{N}=0.06$  & 14& 0.078&5.7\\
$\sigma_{N}=0.08$  & 15& 0.103&7.5\\
$\sigma_{N}=0.1$   & 20& 0.129 &9.4\\
\hline
\end{tabular}
\label{tab:rmse_results}
\end{table}
\end{center}
\begin{figure}[t!!!!!]
\begin{center}
\includegraphics[trim= 0cm 1.2cm 0cm 0.5cm,clip=true,width= 7.8 cm]{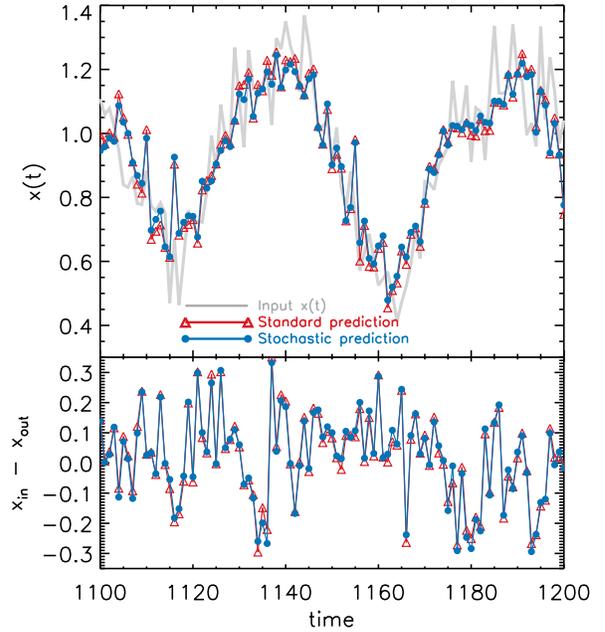}
\caption{Predictions of Mackey--Glass noisy chaotic time series with a white noise contribution of $\sigma_N=0.1$. The grey solid line correspond to the original Mackey--Glass noisy chaotic time series.
The red and blue lines identified the results from {\it standard} {\sc ANN+PSO} and {\it stochastic} {\sc ANN+PSO}, respectively.
The upper panel draw the $y_{i}$ and $\hat{y}_{i}$ predictions, and the lower panel the residual contribution ($x_{\rm in} - x_{\rm out}$) of both methods.
}
\label{fig:MG_output_0.1_prediction}
\end{center}
\end{figure}
\begin{figure}[t!!!!!]
\begin{center}
\includegraphics[width= 7.5 cm]{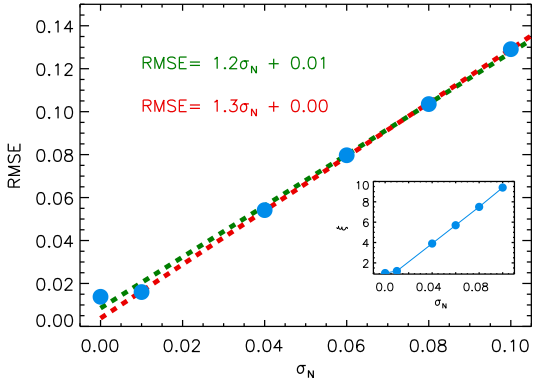}
\caption{Impact of the noise on the performance prediction.}
\label{fig:MG_output_rmse_vs_sigma}
\end{center}
\end{figure}

{\it The prediction performance.}
As illustration, the predictions obtained for noisy case $\sigma_{N}=0.1$, from the {\it standard} {\sc ANN+PSO} ($y_i$) and the {\it stochastic} {\sc ANN+PSO} ($\hat{y}_i$) procedures, are presented in Figure~\ref{fig:MG_output_0.1_prediction}.
As expected, even on this high noise level case, the $y_{i}$ and $\hat{y}_{i}$ predictions are in total agreement.
Actually, the RMSE obtained from both methods is same (in the approximation of the third decimal) for each noisy case. For this reason, the RMSE shown in Table~4 represent the RMSE of both methods.

On the other hand, and as expected, the RMSE increases as growing the noise level (see Figure~\ref{fig:MG_output_rmse_vs_sigma}).
For example, we obtained RMSE of 0.0138 and 0.13 for the noiseless and noisy (with $\sigma_{N}$\,=\,0.1) cases, respectively.
From Figure~\ref{fig:MG_output_rmse_vs_sigma}, we observe a linear correlation between the RMSE and the input noise level. The best fit model, without considering the RMSE of the noiseless case, corresponds to RMSE\,$=\,1.3\,\sigma_{N}$, which shows a strong lineal correlation.
Therefore, we confirm that a higher noise level in input data leads to a poor estimation of the prediction estimator, which is related linearly with the input noise level.

Also, the ratio $\xi=$\,RMSE$_{\rm noisy}$/RMSE$_{\rm noiseless}$ (third column in Table~\ref{tab:rmse_results}) can be used to study the impact of noise on the performance efficiency of our implementation (with respect to nominal case).
The bottom-right panel of Figure~\ref{fig:MG_output_rmse_vs_sigma} shows the performance efficiency  against the input noise level.
In the worst case, the performance efficiency ($\xi$) is strongly affected by  one order of magnitude with respect to noiseless case. Even so, the {\it standard} and {\it stochastic} ANN+PSO confirm to be a powerful tools for making predictions of chaotic time series.

In the literature, we do not find a similar implementation (due to the ahead prediction, type and level of noise, etc.) that allows us a straightforward comparison of results.
For example, we can contrast our results with those presented by Sheng et al. 2012 \cite{sheng2012}. They applied the Echo State Network based on dual estimation ({\sc ESN}) on a noisy Mackey--Glass time series (with a sampling of 2 second) with a white noise level of $\sigma=0.1$.
However, the ahead prediction was one, which is considerable lower than ours.
Yet let us carry out a plain comparison.
Depend on the prediction performance, they obtained RMSE of 0.05 for {\sc Generic ESN} (hereafter {\sc GESN}) and 0.04 for {\sc CKF/KF based ESN} (henceforth {\sc CESN}).
In this context, the impact of the noise on the performance efficiency is lower in ANN+PSO implementation (with respect to the ESN).
In fact, we have a performance efficiency $\xi$ of 9.4 while they obtained $\xi$ of 1161 and 33.5 for {\sc GESN} and {\sc CESN}, respectively.

\begin{figure}[t!!!!!]
\begin{center}
\includegraphics[trim= 0cm 1.2cm 0cm 0.cm,clip=true,width= 7.9 cm]{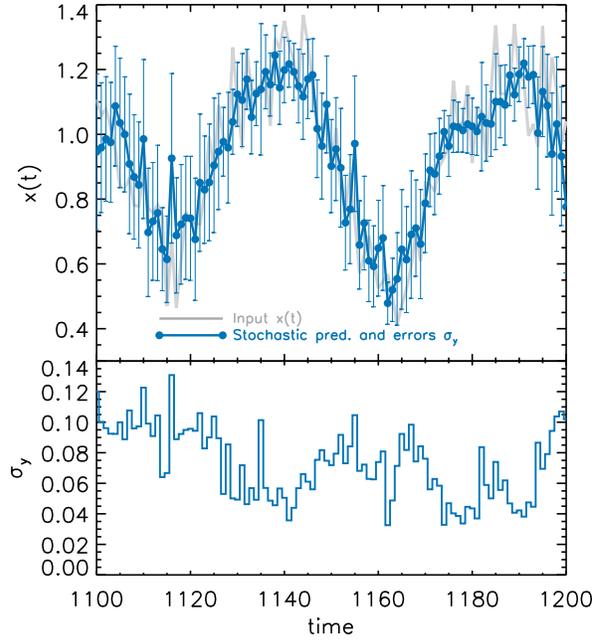}
\caption{Predictions and uncertainties from the {\it stochastic} {\sc ANN+PSO} for the Mackey--Glass chaotic time series. This corresponds to case with a white noise of $\sigma_N=0.1$. The grey solid line draws the original Mackey--Glass noisy chaotic time series. The blue points with error bars correspond to the $\hat{y}_{i}$ prediction and their uncertainties $\sigma_{\hat{y}}$. For optimal display of the uncertainties, these are presented in the low panel.}
\label{fig:MG_output_0.1}
\end{center}
\end{figure}
{\it Prediction uncertainties.}
One of a main goals of this work is to estimate the uncertainty on the prediction.  The prediction measurement ($\hat{y}_i$) and the error bars ($\sigma_{\hat{y}_i}$) obtained from the {\it stochastic} {\sc ANN+PSO}, for the noisy time series with $\sigma_{N}=0.1$, are presented in Figure~\ref{fig:MG_output_0.1}. 
We confirm that our forecast and the input data, for the strong noise contribution, are in agreement at one-sigma (at 68.5\% of confidential level) when the error bars are considered.
The uncertainties obtained are presented in the low panel of Figure~\ref{fig:MG_output_0.1}. 
We found a maximum and minimum uncertainty of 0.024 and 0.13, respectively, with a average of $<\sigma_{\hat{y}_i}>=0.07$.
This value is lower than the input noise level ($ <\sigma_{\hat{y}_i}> / \sigma_{N}= 0.7$), and this show impact of the error propagation in our methods.
According to Figure~\ref{fig:MG_output_0.1}, a relationship between the uncertainties and the times is not appreciated. 

Finally, and from Figures~\ref{fig:MG_output_0.1_prediction}~and~\ref{fig:MG_output_0.1}, we have proven that the ANN+PSO (with the {\it standard} and/or the {\it stochastic} implementation) is a robust tool in the predictability (for the short-term prediction) of time series affected by a white noise.
In addition, now the ANN+PSO method can provide, for first time, an estimation of the uncertainty of the prediction.

\section{Conclusions}
\label{sec:conclusion}
In this paper, a hybrid algorithm based on artificial neural network and particle swarm optimization (ANN+PSO) is used in the short-term $x(t+6)$ prediction of Mackey--Glass chaotic time series.
In addition, an study of the impact of the noise on our hybrid method is presented.
Based on the results and discussion presented in this study, we have the following conclusions:

\noindent$\bullet$ The current value $x(t)$ and the past values used have influential effects on the good training and predicting capabilities of the chosen network.

\noindent$\bullet$ In noiseless case, simulation shows that this hybrid ANN+PSO algorithm is a very powerful tool for making prediction of chaotic time series, and the low deviations found with the proposed method show an accuracy comparable with other methods available in the literature.

\noindent$\bullet$ In noisy cases, we have proven that the hybrid {\sc ANN+PSO} is a robust tool in the predictability of the short-term prediction of chaotic time series affected by a white noise.

\noindent$\bullet$ The impact of the noise on the topology and performance efficient of the ANN+PSO is important.
However, this study shows that the error propagation through the ANN+PSO have a linear behaviour, which generates a linear relationship between the RMSE (optimization parameter) and the input noise level. Therefore, the PSO optimization provides a linearity which ensures that the neural network will converge to an appropriate solution, even if a noise level contribution is present.

\noindent$\bullet$ For noisy cases, although a straightforward comparison with literature is unavailable. The performance efficient $\xi$ proves that the {\it standard}/{\it stochastic} ANN+PSO implementation is affected in a lesser degree than other similar performance.

\section*{Conflict of Interests}
The authors declare that there is no conflict of interests
regarding the publication of this paper.

\section*{Acknowlegments}
The authors thank support from the Research Directorship of the University of La Serena (DIULS).

\end{document}